\documentclass{article}
\usepackage{spconf,amsmath,graphicx,hyperref}
\usepackage{multirow,booktabs,adjustbox,subfig}
\usepackage{algorithm,algpseudocode}

\title{Relative Mismatch: Local-Reference Calibration of Feature-Space Flows
for Anomalous Sound Detection}

\makeatletter
\def\@name{%
  \emph{Anbai Jiang$^1$, Xinhu Zheng$^2$, Lvxin Xu$^1$, Shuwei Zhang$^1$, Wenrui Liang$^1$}\\
  \emph{Pingyi Fan$^{1*}$, Wei-Qiang Zhang$^1$, Cheng Lu$^3$, Jia Liu$^{1,4}$}%
}
\makeatother
\address{
$^1$Tsinghua University \quad
$^2$Shanghai Jiao Tong University \\
$^3$North China Electric Power University \quad
$^4$Huakong AI Plus
\thanks{*Corresponding Author\\ \indent \indent
This work was supported by the National Key Research and Development Program of China
(Grant No. 2021YFA1000500(4)) and the National Natural Science Foundation of China
under Grant No. 62276153.}
}

\graphicspath{{imgs/}}

\begin{document}
\maketitle

\begin{abstract}
Anomalous sound detection (ASD) has long been dominated by $k$-nearest-neighbor (KNN) based detectors, which essentially perform implicit likelihood estimation over normal samples. In this work, we investigate whether generative models can better serve this role. We propose Relative Mismatch, a generative ASD backend powered by flow matching, which learns a velocity field that transports Gaussian noise to a representative feature space of normality. During inference, it measures the mismatch between the oracle and predicted path velocities and aggregates them through a two-level design. To mitigate the inherent mismatch offsets incurred by domain shift, each query is further calibrated with the mismatch of its local normal reference, thereby exposing only its deviation beyond normality. Extensive experiments on DCASE 2020--2025 demonstrate that Relative Mismatch outperforms state-of-the-art backends with the highest score of 71.01, along with strong robustness and training stability. Furthermore, we show that curating a compact and discriminative feature space is the key to unleash the power of generative models for ASD.
\end{abstract}


\section{Introduction}
\label{sec:intro}

Anomalous sound detection (ASD) identifies malfunctioning machine sounds using only normal recordings for training. The state-of-the-art (SOTA) paradigm~\cite{jiang2024anopatch,jiang2025adaptive} employs foundation-model-based frontends for feature extraction and statistical anomaly detection (AD) backends, where $k$-nearest-neighbor (KNN) based detectors~\cite{wilkinghoff2025local,matsumoto2025adjusting,wilkinghoff2026mind} are generally the best. As is known, KNN is a non-parametric likelihood estimator, while generative models are essentially parametric ones. Therefore, can a generative model learn a more powerful notion of normality than KNN provides?

In fact, flow matching has recently shown promise in AD of other modalities. Flow Matching~\cite{lipman2022flow} learns a time-dependent vector field to transport samples from a simple prior to the target distribution. When trained on normal data, the discrepancy between the oracle and the predicted velocities reveals the degree of abnormality. TCCM~\cite{li2026scalable} narrows the prior as the Dirac delta at the origin and simplifies the flow as the contraction vector to the origin. Flow Mismatching~\cite{chen2026flow} combines the velocity mismatch at different time steps with different starting points. These works demonstrate the AD potential of learned generative dynamics. However, Flow Mismatching still operates in the sparse and high-dimensional sample space, leaving it lagging behind feature-space KNN-based AD~\cite{roth2022towards}. Thus, we argue that sample-space modelling devotes capacity to abundant variation irrelevant to anomaly, obscuring the deviations that actually matter for AD.

In this paper, we propose \textbf{Relative Mismatch} as a new generation ASD backend. We first transfer Flow Mismatching to the compact, task-specific audio feature space, removing AD-irrelevant structure and concentrating the model on the most discriminative part. We then propose local-reference calibration to improve the ASD with domain shift, where data of uncommon yet normal working conditions (target domain) are far scarcer than data of common conditions (source domain). This biases the model and creates condition-dependent offsets for scarce conditions that interfere with anomaly ranking. To reduce these offsets, we measure the relative mismatch, which reveals how strong a query mismatch deviates from its local normality. The proposed scheme is evaluated rigorously across 6 DCASE ASD datasets~\cite{koizumi2020description,kawaguchi2021description,dohi2022description,dohi2023description,Nishida_arXiv2024_01,nishida2025description}, where it generally outperforms multiple SOTA backends, showing remarkable diagnostic accuracy and trustworthy robustness. To the best of our knowledge, Relative Mismatch is the first generative model that surpasses KNN backends. Therefore, our contributions are summarized as follows:
\begin{itemize}
\item We propose Relative Mismatch, the first generative-model-based ASD backend that generally outperforms SOTA backends with trustworthy robustness.
\item We demonstrate that modeling space curation is the key for generative-model-based ASD backend.
\item We introduce local-reference calibration to reduce the mismatch offsets on minor working conditions.
\end{itemize}

\section{Relative Mismatch}
\label{sec:method}

\begin{figure}[t]
  \centering
  \includegraphics[width=0.98\linewidth]{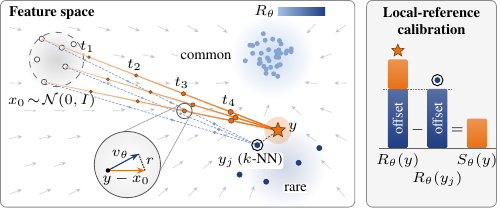}
  \caption{Detection mechanism of Relative Mismatch. Left: velocity mismatches of a query $y$ along Gaussian paths at $\{t_1,\dots,t_T\}$ are aggregated into $R_\theta(y)$. Right: the mismatch of the nearest training feature $y_j$ is subtracted as a local reference to reduce offsets incurred by domain shift.}
  \label{fig:arch}
\end{figure}

In this section, the proposed scheme is elaborated in three progressive aspects: how to train the flow matching model in the feature space, how to leverage its flow mismatch to detect anomalies, and how local-reference calibration is employed.

\subsection{Flow Matching Training in Feature Space}
\label{ssec:features}

We train a Rectified Flow~\cite{liu2023flow} model $v_\theta(x_t,t)$ in the compact feature space to learn normal audio dynamics. The feature space is derived by fine-tuning audio foundation models via working condition classification~\cite{jiang2024anopatch,jiang2025adaptive}. Given a set of $D$-dimensional normal training features $\mathcal{Y}$, we independently sample a feature $y\in\mathcal{Y}$, a Gaussian starting point $x_0\sim\mathcal{N}(0,I_D)$, and a time step $t\sim\mathcal{U}(0,1)$. The linear interpolation path from $x_0$ to $y$ is:
\begin{equation}
 x_t=(1-t)x_0+ty.
 \label{eq:path}
\end{equation}
Its conditional path velocity is:
\[
 u(x_0,y)=\frac{\mathrm{d}x_t}{\mathrm{d}t}=y-x_0.
\]
Thus, $y-x_0$ is the oracle target velocity for the sampled endpoint pair $(x_0,y)$. The model $v_\theta(x_t,t)$ learns to predict this conditional target from the interpolated feature and time, whose training objective is:
\begin{equation}
 \mathcal{L}(\theta)=\mathrm{E}_{y,x_0,t}
 \left[\left\|v_\theta(x_t,t)-(y-x_0)\right\|_2^2\right].
 \label{eq:loss}
\end{equation}
Sampling $(y,x_0,t)$ at each optimization step trains the model on oracle conditional velocities from normal endpoint pairs.

\begin{figure}[t]
  \centering
  \subfloat[Absolute mismatch]{
    \includegraphics[width=0.47\linewidth]{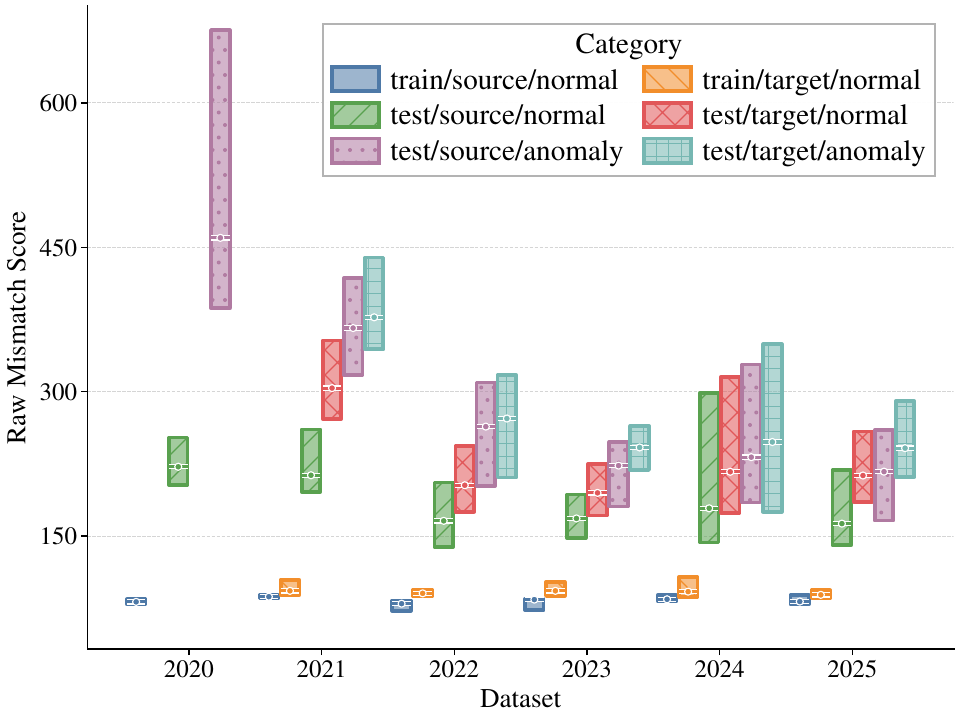}
    \label{fig:raw_mismatch}
  }
  \hfill
  \subfloat[Relative mismatch]{
    \includegraphics[width=0.47\linewidth]{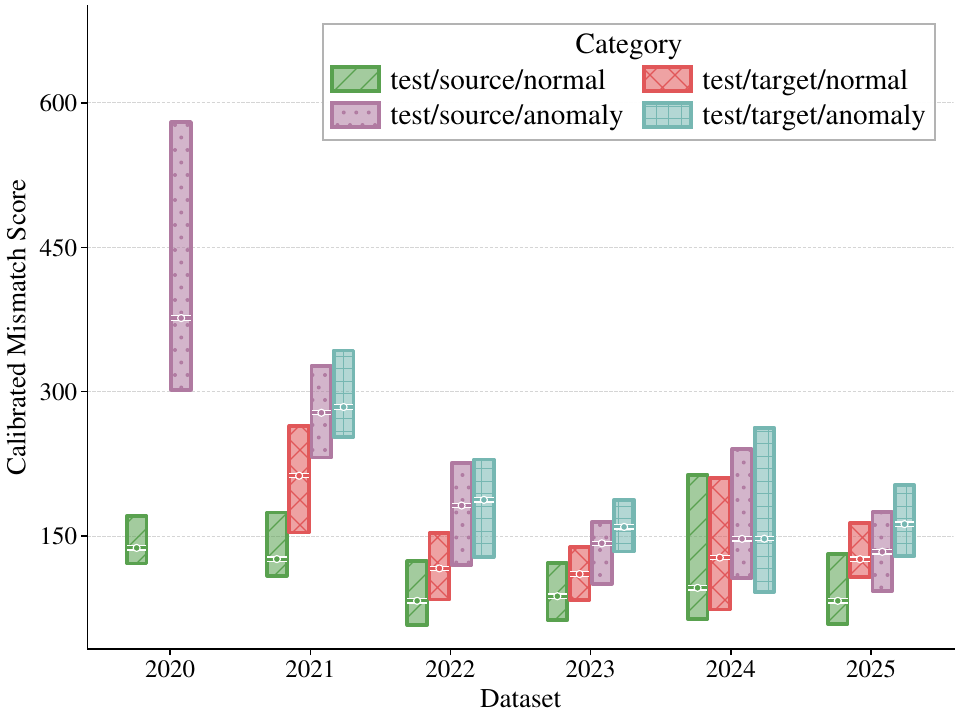}
    \label{fig:calib_mismatch}
  }
  \caption{Absolute flow mismatch and relative mismatch scores across DCASE datasets. Markers and boxes show the median and 25\%--75\% range of mismatch scores. Target-domain samples have higher mismatch than their source-domain counterparts (2021--2025), while local-reference calibration reduces this gap and make them more comparable.}
  \label{fig:mismatch_value}
\end{figure}

\subsection{Inference with Absolute Flow Mismatch}
\label{ssec:absolute}

Trained on normal embeddings, $v_\theta(x_t,t)$ captures the normal transport dynamics, whose discrepancy with the target velocity of a test endpoint $y$ is likely to indicate a potential anomaly during inference. Therefore, we detect anomalies via flow mismatches at multiple time steps of multiple trajectories following~\cite{chen2026flow}. Specifically, the flow mismatch at time step $t$ with a trajectory starting from $x_0$ is measured as:
\begin{equation}
 r_\theta(y,x_0,t)=\|v_\theta(x_t,t)-(y-x_0)\|_2^2,
 \label{eq:residual}
\end{equation}
which is essentially the same with the training objective, capturing disagreement in both direction and magnitude.

Flow mismatch is measured on multiple trajectories with different time steps and starting points. We draw $T$ observations $\{t_1,t_2,...,t_T\}$ along time with equal slicing: $t_i=i/(T+1)$. For each $t_i$, we sample $K$ unique Gaussian starting points, resulting in a Gaussian bank $\{x_0^{(m,i)}\}_{m=1,i=1}^{K,T}$. This bank is fixed and reused for every scored training and test embedding. We quantify the flow mismatch of each $(x_0^{(m,i)},t_i)$ via Eq.~\ref{eq:residual}, and these scores $\{r_\theta(y,x_0^{(m,i)},t_i)\}_{m=1,i=1}^{K,T}$ are then aggregated through a two-level hierarchy. For each fixed time step $t_i$, the mismatch over the $K$ Gaussian paths is first aggregated via mean pooling:
\begin{equation}
  \bar r_\theta(y,t_i)=\frac{1}{K}\sum_{m=1}^{K}
  r_\theta\!\left(y,x_0^{(m,i)},t_i\right).
  \label{eq:path-aggregation}
\end{equation}
We then aggregate this profile over the $T$ interior time steps:
\begin{equation}
 R_\theta(y)=\sum_{i=1}^{T}w_i\,\bar r_\theta(y,t_i),
 \label{eq:absolute}
\end{equation}
where $w_i=t_i^2/\sum_{\ell=1}^{T}t_\ell^2$ is the quadratic time-step weight following Flow Mismatching, emphasizing positions closer to the feature endpoint. Thus, time aggregation converts the path-aggregated profile into one scalar score: larger $R_\theta(y)$ indicates stronger disagreement with normal dynamics and hence greater evidence of an anomaly. Scoring requires only forward evaluations at the constructed interpolation points; no sample reconstruction or numerical integration of a generative trajectory is needed.

\begin{algorithm*}[t]
\caption{Relative Mismatch inference with local-reference calibration}
\label{alg:inference}
\begin{algorithmic}[1]
\Require Trained $v_\theta(x_t,t)$; Training features $\mathcal{Y}$;
  $K,T,k$
\State Sample a shared $K\times T$ bank of independent
  $x_0^{(m,i)}\sim\mathcal{N}(0,I_D)$
\For{each query feature $y$}
  \State Find its $k$ Euclidean nearest neighbors
    $\mathcal{N}_k(y)$ in $\mathcal{Y}$
  \For{each time step $t_i=i/(T+1)$, $i=1,\ldots,T$}
    \For{each path $m=1,\ldots,K$ with starting point $x_0^{(m,i)}$}
      \State Compute query mismatch $r_{m,i}(y)$ and neighbor
        mismatches $\{r_{m,i}(y_j)\}_{j\in\mathcal{N}_k(y)}$
        using Eq.~\ref{eq:residual}
      \State Relative mismatch:
        $\widetilde r_{m,i}(y)\gets r_{m,i}(y)
        -\frac{1}{k}\sum_{j\in\mathcal{N}_k(y)}r_{m,i}(y_j)$
        \Comment{$\widetilde{\cdot}$ denotes calibrated counterpart, cf.\ Eq.~\ref{eq:relative}}
    \EndFor
    \State Path aggregation:
      $\overline{\widetilde r}_i(y)\gets\frac{1}{K}\sum_{m=1}^{K}\widetilde r_{m,i}(y)$
      \Comment{$\overline{\cdot}$ denotes scores after path aggregation, cf.\ Eq.~\ref{eq:path-aggregation}}
  \EndFor
  \State Time aggregation:
    $S_\theta(y)\gets\sum_{i=1}^{T}w_i\overline{\widetilde r}_i(y)$,
    where $w_i=t_i^2/\sum_{\ell=1}^{T}t_\ell^2$
  \State Output $S_\theta(y)$ as the final ASD score
\EndFor
\end{algorithmic}
\end{algorithm*}

\subsection{Local-Reference Calibration}
\label{ssec:relative}

A key assumption of Flow Mismatching is: the model should be well-trained on the normal data so that the learned flow velocity approximates the oracle marginal velocity, and thus the irreducible component of the mismatch is approximately constant across samples. However, there is severe domain imbalance in DCASE datasets, where the training data in the target domain (minor working conditions) is far less than that of the source domain (common working conditions). This incurs the model to be under-trained on the target domain, leading to inherent mismatch offsets for the target domain. As shown in Fig.~\ref{fig:raw_mismatch}, the absolute mismatch scores of the target domain are notably larger than their source domain counterparts, which contradicts the DCASE criteria that anomalies should be detected under the same threshold for both domains.

To this end, we employ local-reference calibration to reduce these offsets, as presented in Fig.~\ref{fig:arch}. Since nearby normal features in the training set have similar working conditions with the query, their mismatch scores serve as an estimate of the local mismatch offset for the query condition, which can be used for calibration. To be specific, let $\mathcal{N}_k(y)$ contain the indices of the $k$-nearest training features in $\mathcal{Y}$ under Euclidean distance. The local mismatch offset can be estimated by:
\begin{equation}
  \widehat b_\theta(y)=\frac{1}{k}\sum_{j\in\mathcal{N}_k(y)}R_\theta(y_j),
  \label{eq:mismatch_ref}
\end{equation}
where $R_\theta(y_j)$ is calculated in the same manner as Eq.~\ref{eq:absolute}, grounded on the same Gaussian sample bank $\{x_0^{(m,i)}\}$ as $R_\theta(y)$. This estimation is essentially the mean mismatch of query neighbors in the training set. The flow mismatch therefore can be calibrated by subtracting the local-reference mismatch from the query mismatch:
\begin{equation}
  S_\theta(y)=R_\theta(y)-\widehat b_\theta(y),
  \label{eq:relative}
\end{equation}
which highlights the deviation of a query mismatch from its normal offset rather than its absolute value. This calibration design further improves the ASD performances on DCASE datasets with domain shifts, as presented in Table~\ref{tab:main}.

Algorithm~\ref{alg:inference} summarizes the complete inference pipeline. For each query feature, the method first retrieves nearby normal training features, then evaluates the query and its local references on the same Gaussian bank at every interior time and path. Their mismatch values are subtracted path by path, averaged over paths, and finally combined across time with the normalized quadratic weights to produce the ASD score. Here $r_{m,i}$ denotes the single-path, single-time mismatch and $R_\theta$ denotes its path--time aggregate; the pre-aggregation subtraction therefore gives the same result as Eq.~\ref{eq:relative} by linearity.

\section{Experiments and Analysis}
\label{sec:exp}

\subsection{Experiment Setup}


The experiment is conducted on DCASE 2020--2025
~\cite{koizumi2020description,kawaguchi2021description,dohi2022description,dohi2023description,Nishida_arXiv2024_01,nishida2025description} (both dev and eval) to rigorously evaluate both the efficacy and the robustness. We first fine-tune BEATs~\cite{chen2022beats} on each dataset separately ($D=128$) with DCASE 2020--2023 using AnoPatch~\cite{jiang2024anopatch} and DCASE 2024--2025 using Adaptive Prototype Learning~\cite{jiang2025adaptive}. We report the official score (\%) for each dataset (hmean over dev and eval) along with an overall arithmetic mean across datasets.

\subsection{Implementation Details}
\label{ssec:setup}

Relative Mismatch employs a residual dense network with 3 blocks, hidden width 512, a 64-dimensional sinusoidal time embedding, LayerNorm, SiLU, and dropout 0.05. This lightweight backend possesses negligible training cost compared to the frontend feature extractor. The model is trained separately for each section. For each section, embeddings are first standardized dimension-wise using the training set statistics. We use AdamW for 4000 optimization updates with batch size 256, learning rate $2\times10^{-4}$, weight decay $10^{-4}$, and gradient-norm clipping at 1.0. The learning rate follows cosine annealing. During inference, we use the interior uniform grid with $K=40$, $T=4$ along with $k=1$. Relative mismatch is experimented under 5 different seeds.

\subsection{Baselines}
\label{ssec:results}

We compare 8 ASD backends on the same set of features. Five KNN-based detectors~\cite{jiang2024anopatch,wilkinghoff2025local,matsumoto2025adjusting,wilkinghoff2026mind} are compared, all of which use cosine distance with top-1 neighbor (best config for ASD). For KNN VarMin~\cite{matsumoto2025adjusting}, we use the TrainAll variant as it works best on our embeddings. Three other commonly used backends are also included: LOF~\cite{breunig2000lof}, Mahalanobis-distance-based detector, and Gaussian mixture model (GMM).

\begin{table}[t]
\centering
\caption{Comparison of ASD backends (↑)}
\label{tab:main}
\adjustbox{max width=\linewidth}{
\begin{tabular}{lrrrrrrr}
\toprule
Backend & 2020 & 2021 & 2022 & 2023 & 2024 & 2025 & Mean \\
\midrule
KNN & 92.58 & 70.53 & 64.32 & 66.11 & 61.97 & 57.31 & 68.80 \\
KNN min~\cite{jiang2024anopatch,jiang2025adaptive} & 92.58 & 70.53 & 66.71 & \textbf{68.87} & \textbf{64.82} & \textbf{61.22} & 70.79 \\
KNN density~\cite{wilkinghoff2025local} & 92.24 & 70.54 & 64.12 & 62.83 & 62.35 & 60.93 & 68.84 \\
KNN VarMin~\cite{matsumoto2025adjusting}  & 93.05 & \textbf{71.70} & 66.57 & 68.05 & 64.14 & 60.59 & 70.68 \\
KNN Mind the Gap~\cite{wilkinghoff2026mind} & 92.61 & 70.53 & 64.32 & 66.11 & 59.85 & 57.31 & 68.46 \\
LOF~\cite{breunig2000lof} & 92.95 & 69.27 & 62.70 & 63.08 & 59.32 & 58.25 & 67.60 \\
Mahalanobis & 92.66 & 69.93 & 63.96 & 57.92 & 56.24 & 56.58 & 66.22 \\
GMM & 86.14 & 65.82 & 63.83 & 60.84 & 57.89 & 56.66 & 65.20 \\
\midrule
Sample-Space Mismatch & 73.87 & 59.33 & 56.68 & 58.89 & 53.76 & 53.57 & 59.35 \\
Feature-Space Mismatch & 93.38 & 70.97 & 66.84 & 67.83 & 63.90 & 60.25 & 70.53 \\
Relative Mismatch (ours) & \textbf{93.44} & 71.31 & \textbf{67.10} & 68.77 & 64.67 & 60.79 & \textbf{71.01} \\
\bottomrule
\end{tabular}
}
\end{table}

\subsection{Main Results}

Table~\ref{tab:main} compares Relative Mismatch with 8 SOTA backends. Relative Mismatch obtains the highest six-year mean of 71.01, exceeding all KNN variants which have been dominating the field. Our method consistently remains in the top tier across datasets without exhibiting substantial degradation on any particular one. Such consistent performance across datasets demonstrates the strong robustness of the proposed scheme, making it a reliable solution for practical use.

\subsection{Core Design Analysis}
\label{ssec:core_designs}

We further analyze the efficacy of two core designs: feature-space modeling and local-reference calibration. For sample-space mismatch, we reproduce vanilla Flow Mismatching on log-mel spectrograms. As presented in the last three rows of Table~\ref{tab:main}, moving from sample space to feature space yields a significant improvement from 59.35 to 70.53, suggesting that the modeling space must be sufficiently compact and highly tailored to AD so as to fully unlock the potential of generative models. On the other hand, applying local-reference calibration further boosts the score to 71.01, with uniform improvements on domain shift datasets (DCASE 2021--2025). We also compare the score distributions of absolute and relative mismatch in Fig.~\ref{fig:mismatch_value}, showing that local-reference calibration effectively narrows the domain gap.


\subsection{Ablation Studies}
\label{ssec:ablation}

\begin{figure}[t]
  \centering
  \subfloat[Selection of $K$ and $T$]{
    \includegraphics[width=0.98\linewidth]{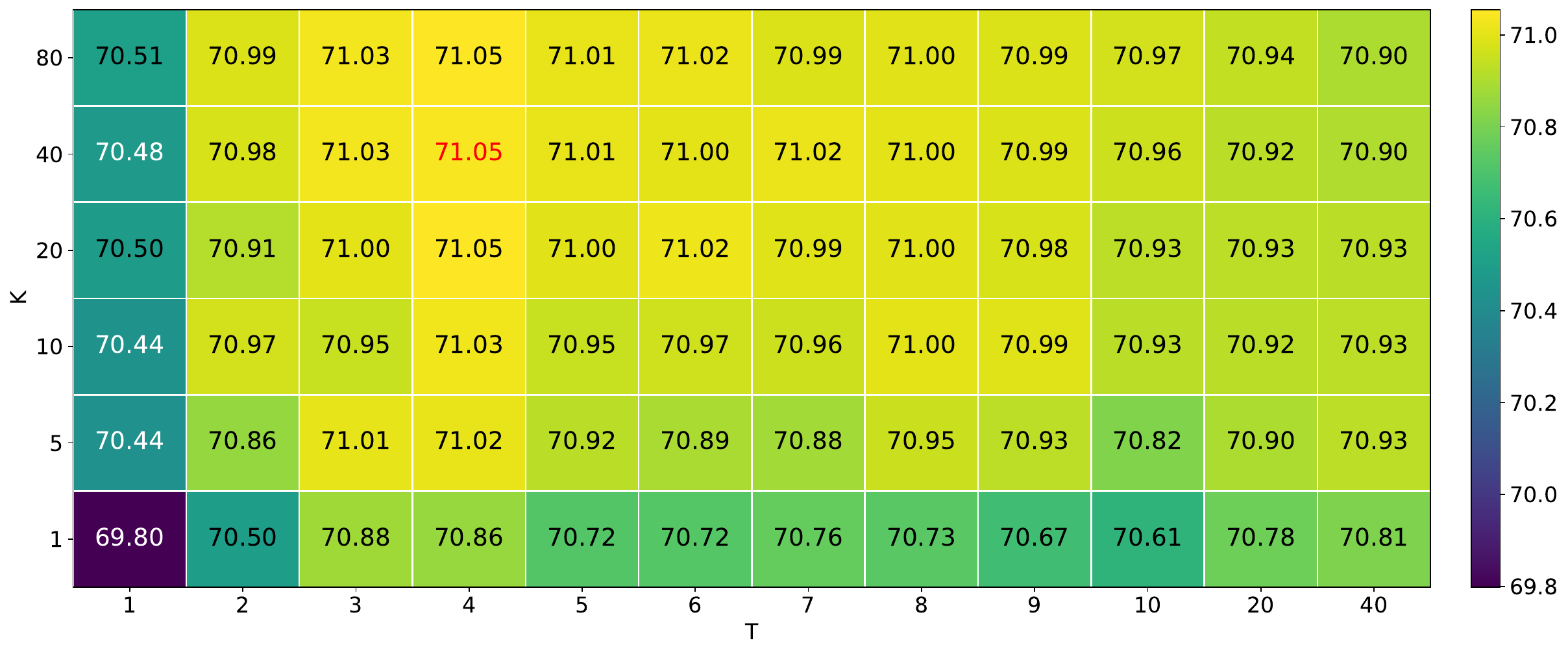}
    \label{fig:ablation_kt}
  }

  \subfloat[Selection of $k$]{
    \includegraphics[width=0.46\linewidth]{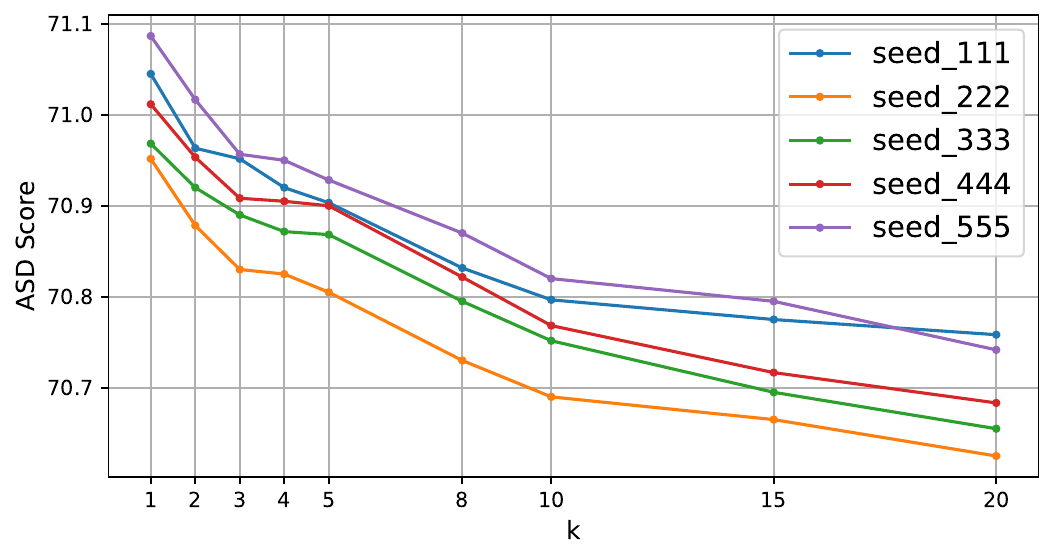}
    \label{fig:ablation_knn_k}
  }
  \hfill
  \subfloat[Training stability]{
    \includegraphics[width=0.46\linewidth]{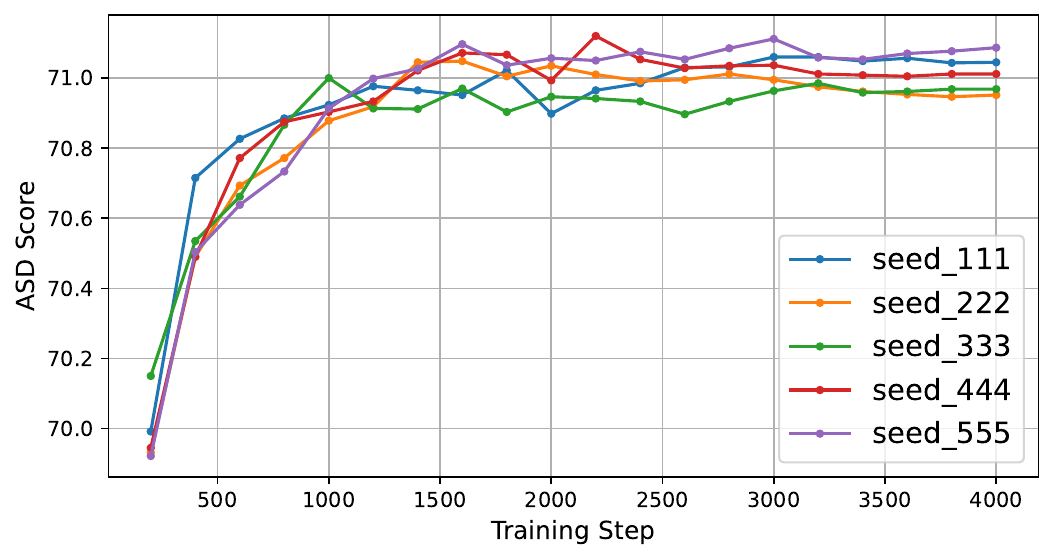}
    \label{fig:ablation_ep_curve}
  }
  \caption{Ablation studies on $K$, $T$, $k$ and training stability.}
  \label{fig:ablation}
\end{figure}

Fig.~\ref{fig:ablation_kt} experiments with different $K$ and $T$ (1 seed), where the performance is consistent within a local region centered at $K=40$ and $T=4$. Fig.~\ref{fig:ablation_knn_k} experiments with different $k$, where the performance drops monotonically as $k$ grows. This suggests that only the nearest neighbor provides reliable offset estimation for the query condition. Fig.~\ref{fig:ablation_ep_curve} presents the performance curves during training, where the performances across seeds are stable, and all curves converge to their best horizons in the latter half of training. Therefore, Relative Mismatch is not sensitive to overfitting, thus making it a robust detector comparable with KNN.

\section{Conclusion}
\label{sec:conclusion}

We proposed Relative Mismatch, the first generative-model-based ASD backend that outperforms SOTA detectors. By curating a compact and discriminative feature space, aggregating velocity mismatches across multiple time steps and trajectories, and calibrating inherent offsets via local-reference subtraction, it unlocks the power of flow matching for typicality estimation. Extensive experiments across 6 DCASE datasets demonstrate superior efficacy, robustness, and training stability, establishing a new paradigm for generative ASD.



\bibliographystyle{IEEEbib}
\bibliography{strings,refs}
\end{document}